\documentclass[10pt]{article}
\usepackage{colordvi,amsfonts,amsmath,epsfig,subfigure,cite}

\usepackage{color}

\numberwithin{equation}{section} 

\newcommand{\be}{\begin{equation}}
\newcommand{\ee}{\end{equation}}
\newcommand{\ba}{\begin{eqnarray}}
\newcommand{\ea}{\end{eqnarray}}
\newcommand{\bi}{\begin{itemize}}
\newcommand{\ei}{\end{itemize}}

\newtheorem{thm}{Theorem}

\newtheorem{rem}[thm]{Remark}
\begin{document}

\title{A Computational Study of Residual KPP Front Speeds in Time-Periodic Cellular Flows 
in the Small Diffusion Limit}
\author{Penghe Zu, Long Chen, Jack Xin \thanks{Department of Mathematics, 
UC Irvine, Irvine, CA 92697, USA.\newline Corresponding author: Penghe Zu. Phone number: 9493784420.\newline \hspace{.2 in} Emails: (pzu,chenlong,jxin)@math.uci.edu.}}
\date{}
\maketitle
\begin{abstract}
The minimal speeds ($c^*$) of the Kolmogorov-Petrovsky-Piskunov (KPP) fronts  
at small diffusion ($\epsilon \ll 1$) in a class of time-periodic cellular flows with chaotic streamlines is investigated in this paper. 
The variational principle of $c^*$ reduces the computation to 
that of a principal eigenvalue problem on a periodic domain of 
a linear advection-diffusion operator with space-time periodic coefficients and small diffusion.
To solve the advection dominated time-dependent eigenvalue problem efficiently over large time, 
a combination of finite element and spectral methods, as well as the associated 
fast solvers, are utilized to accelerate computation. 
In contrast to the scaling $c^*=\mathcal{O}(\epsilon^{1/4})$ in steady cellular flows, 
a new relation $c^* = \mathcal{O}(1)$ as $\epsilon \ll 1$ is revealed in the time-periodic cellular flows due to the presence of chaotic streamlines. 
Residual propagation speed emerges from the Lagrangian chaos which is quantified as a sub-diffusion process.
\end{abstract}
\medskip

\providecommand{\keywords}[1]{\textbf{\textit{Keywords:}} #1}
\thispagestyle{empty}
\setcounter{equation}{0} \setcounter{page}{0}
\newpage
\section{Introduction}
Front propagation in complex fluid flows arises in many scientific 
areas such as combustion~\cite{R,W85}, population
growth of ecological communities (plankton) in the ocean~\cite{A}, and 
reactive chemical front in liquids~\cite{PS05,X09}. 
An interesting problem is the front speed enhancement 
in time dependent fluid flows with chaotic streamlines and random media~\cite{PS05,X09}. 

In this paper, we shall consider a two dimensional time dependent cellular flow:
\be
(\cos(2\pi y) + \theta \sin(2\pi y)\cos(t), \, \cos(2\pi x) + \theta \sin(2\pi x)\cos(t)), \; \; 
\theta \in (0,1], \label{tdcell}
\ee 
whose steady part $(\cos (2\pi y), \cos(2\pi x))$ is subject to 
time periodic perturbation that causes transition to Lagrangian chaos. 
Chaotic behavior of a similar flow field is qualitatively analyzed by formal dynamical system 
methods in~\cite{CW91}. The enhanced residual diffusion in (\ref{tdcell}) is observed numerically in~\cite{BCVV95}. 
The enhanced diffusion and propagation speeds in steady cellular flows with ordered streamlines and 
their extensions have been extensively studied~\cite{ACVV02,Aud,ChSo,FP,Nov11,H,NR05,RZ07,SZX13,Liu_13,Yu1,Yu2} among others.  
We shall take a statistical look at the 
chaotic streamlines of (\ref{tdcell}) in terms of the scaling of 
the mean square displacements from random initial data, and  
quantify the Lagrangian chaos of (\ref{tdcell}) as a sub-diffusion process. 

Consider then the advection-reaction-diffusion equation:
\begin{equation}\label{eq:nonlinear1}
\partial_t u = \epsilon \triangle u +\vec{B}(x,t)\cdot\nabla u + \frac{1}{\tau}f(u), \quad x\in \mathbb{R}^2, t>0, 
\end{equation}
where $f(u)=u(1-u)$ is the Kolmogorov-Petrovsky-Piskunov (KPP) nonlinearity,
$\epsilon$ is the molecular diffusion parameter,  
$\tau$ is the reaction rate and $\vec{B}(x,t)$ is a space-time periodic, mean zero, and incompressible 
flow field such as (\ref{tdcell}).  If the initial data for $u$ is nonnegative and compactly supported, 
the large time behavior of $u$ is an outward propagating front,  
with speed $c^*=c^*(\vec{e})$ in the direction $\vec{e} = \langle1,0\rangle$. The variational principle of $c^*$ is~\cite{NX05}:
\begin{equation}\label{eq:inf}
c^*=\inf_{\lambda>0} \frac{\mu(\lambda)}{\lambda},
\end{equation}
where $\mu(\lambda)$ is the principal eigenvalue of the periodic-parabolic operator:
\begin{equation}
L^{\lambda}\Phi = \epsilon\triangle\Phi + (\vec{B}+2\lambda \vec{e})\cdot \nabla\Phi + 
(\epsilon \lambda^2 + \lambda \vec{B}\cdot \vec{e} + \tau^{-1}\, f'(0))\Phi -\Phi_t, \label{varp1}
\end{equation}
on the space-time periodic cell $\Omega$.
The principal eigenvalue $\mu(\lambda)$ can be computed by solving the evolution problem:
\ba
&& w_t  =  \epsilon\triangle w +(2\epsilon\lambda\vec{e} + \vec{B}(x,t))\cdot\nabla w + 
 (-C_M +\epsilon\lambda^2 +\lambda\vec{e}\cdot \vec{B}(x,t) + \frac{1}{\tau}f'(0)) w \nonumber \\
&& w(x,0) = 1. \label{para}
\ea
Here the constant $C_M = \epsilon\lambda^2 +\lambda||\vec{B}||_{\infty} 
+ ||\frac{1}{\tau}f'(0)||_{\infty}$, 
so that $w$ remains positive and bounded by one.  The $\mu (\lambda)$ is then given by:
\begin{equation}\label{lyapunov}
\mu(\lambda) = C_M + \lim_{t\rightarrow\infty} \frac{1}{t} \ln\int_{\Omega} w(x,t)\, dx
\end{equation}
The number $\mu(\lambda)$ is also the principle 
Lyapunov exponent of the parabolic equation (\ref{para}), and the formula 
(\ref{lyapunov}) extends to the more general case when $\vec{B}$ is a stationary 
ergodic field\cite{NX09}. The limit then holds almost surely and 
$\mu$ is deterministic\cite{NX09}. KPP fronts are examples of the so called ``pulled fronts''~\cite{S03} 
because their speed is determined by the behavior of the solution far 
beyond the front interface, in the region where the solution is close to zero 
(the unstable equilibrium). The minimal speed of the planar wave solution of the linearized equation 
near the unstable equilibrium gives (\ref{eq:inf}), also known as the marginal stability criterion\cite{S03,X09}.  

Equation (\ref{para}) and formula (\ref{lyapunov}) 
will be discretized for approximating $\mu$ at sufficiently large time for a range of 
$\lambda $ values. The minimal point of $\mu (\lambda)/\lambda $ is searched by the 
golden section algorithm. At each new search of $\lambda$, 
the large time solution of (\ref{para}) is computed. Because $\epsilon$ is small, 
(\ref{para}) is advection dominated. To this end, 
upwinding type finite element methods (EAFE~\cite{XZ99}) and spectral method
(SM) are utilized. Spectral method has high accuracy but the computation is slower 
because it takes small time step due to the restriction of stability. On the other
hand, EAFE can be solved faster using relative larger time step with
implicit discretization. Without the compromise of accuracy, our strategy to obtain 
the minimal point is to narrow down the search interval by EAFE first, and solve
more accurately in a smaller interval by SM.


The rest of the paper is organized as follows. In section 2, we illustrate the 
difference of streamline geometry and dynamical properties of 
steady and unsteady cellular flows. We find that 
the chaotic streamlines of the time periodic cellular flow (\ref{tdcell}) 
can be quantified as a sub-diffusive process with distinct $\theta$-dependent scaling exponents. 
The resulting motion appears ergodic inside the invariant infinite channel domains.
In section 3, we discuss numerical methods for advection-dominated problems 
such as EAFE, and semi-implicit SM, as well as the stability of 
these methods and time step constraints. 
In section 4, we show numerical results of KPP front speeds by 
a combination of these methods and the existence of  
residual front speed in the sense that $c^*(\epsilon) = O(1) > 0$ as $\epsilon \downarrow 0$. In contrast, 
$c^{*} = O(\epsilon^{1/4})$ in steady cellular flows \cite{NR05}. 
The streamlines of the steady cellular flows are ordered with closed orbits 
leading to a much slower rate of enhancement for the transport. We also observe the presence of 
layer and circular structures in the generalized eigenfunction $w$ at large time, a 
reflection of the advection-dominated transport in time periodic cellular flow (\ref{tdcell}). The 
corresponding energy spectrum of $w$ shows decay towards high frequencies with an intermediate scaling range. 
In section 5, we give concluding remarks about our findings.

\section{Properties of Cellular Flows}
 In this section, we illustrate the 
difference of streamline geometry and dynamical properties of 
steady and unsteady cellular flows. In particular, we quantify the chaotic 
behavior of the motion along streamlines of (\ref{tdcell}) using the empirical 
mean square distance inside infinite invariant channels 
in the direction $(1,1)$, or $E[|X(t)\cdot (1,1)|^2]$, where 
$X\in \mathbb R^2$ denotes the particle trajectories in the flow. 
We show computationally that it scales 
with time as $O(t^p)$, $p \in (0,1)$, for sampled values of $\theta \in (0,1]$.
\subsection{Steady Cellular Flow}
\begin{figure}
\includegraphics[width=1 \textwidth, height = 0.8\textwidth]{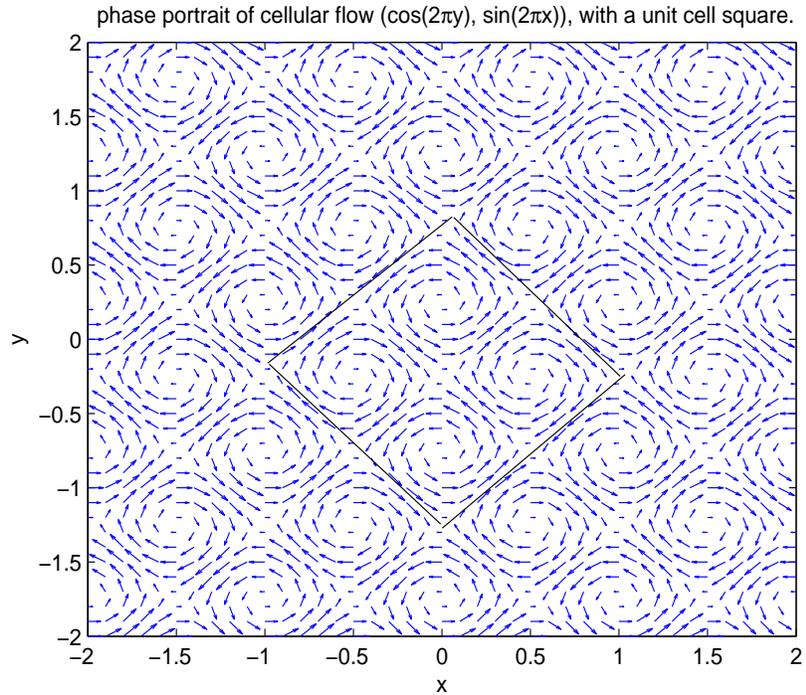}
\caption{Ordered and localized streamlines of steady cellular flow (\ref{steady}) with a cell square containing 
four vertices.}\label{stcell}
\end{figure}

A phase portrait of the steady cellular flow 
\be
B_s=(\cos (2\pi y), \cos (2\pi x)), \label{steady}
\ee
with a cell square is presented in Fig.~\ref{stcell}. 
The phase portrait away from the square simply repeats. The streamlines are ordered. The 
closed orbits form elliptic part of the phase space. The 
saddles are located at half-integer points $(n,m)/2$, $n$,$m \in \mathbb{Z}$ 
with connecting separatrices forming hyperbolic part of the phase space. 
Any particle trajectory is either a closed orbit or a separatrix, hence the motion is 
bounded inside the square.
  
\subsection{Unsteady Cellular Flow}
\begin{figure}[htdp]
\includegraphics[width=0.5\textwidth, height = 0.45\textwidth]{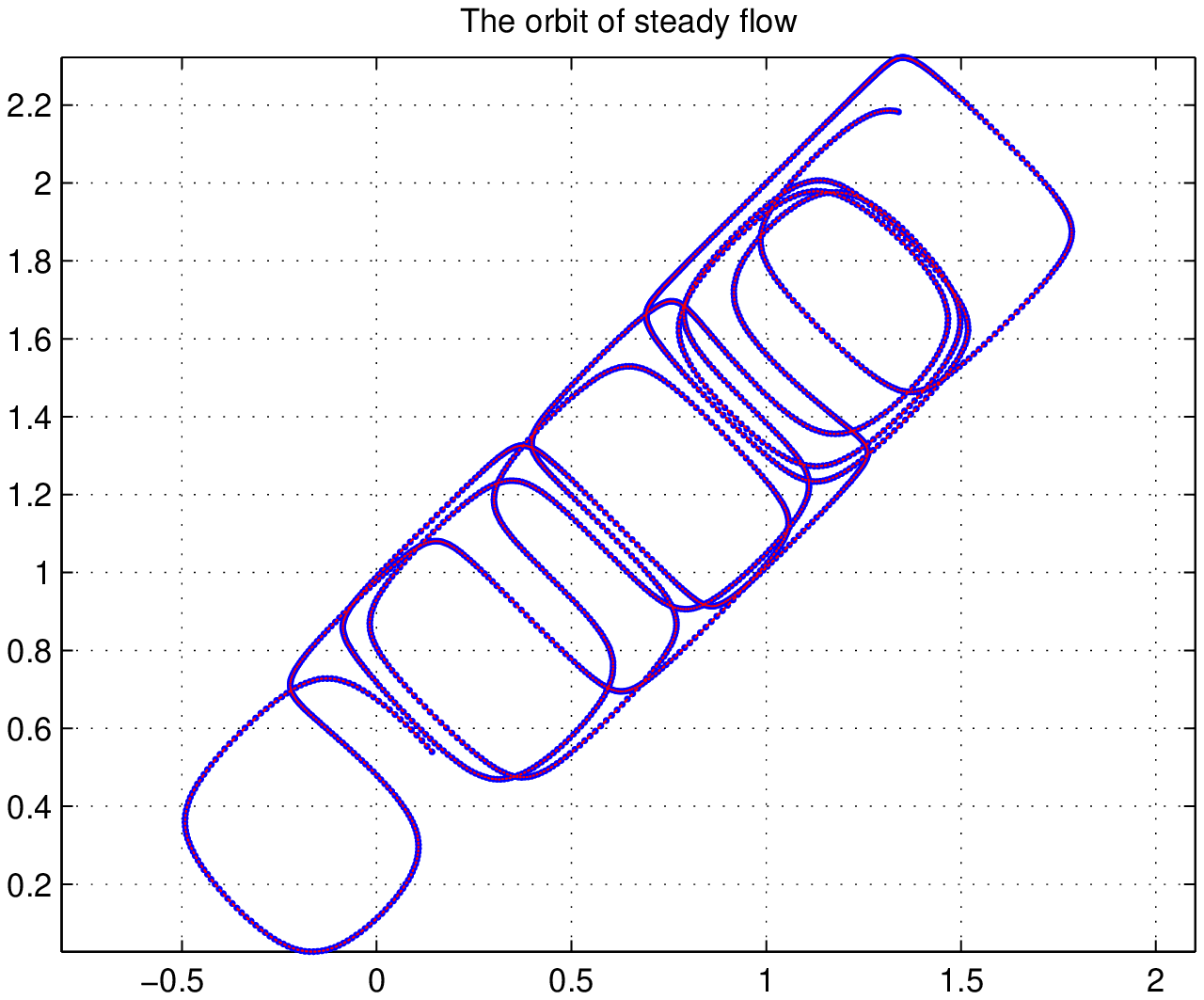}
\includegraphics[width=0.5\textwidth, height = 0.45\textwidth]{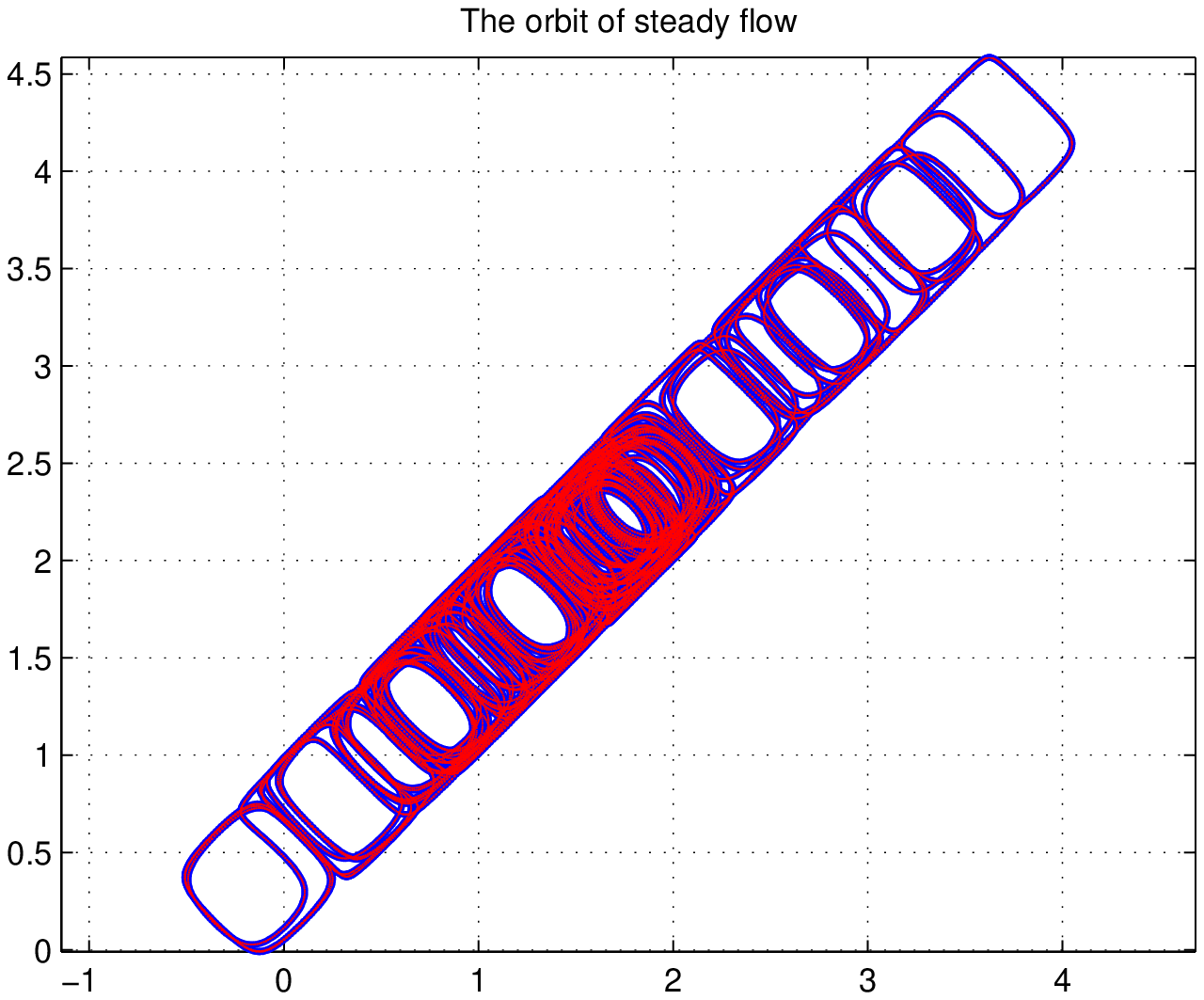}
\caption{Disordered and extended trajectories in time periodic cellular flow (\ref{unsteady})  
over time $[0,20]$ (left) and $[0,200]$ (right) in a channel bounded by invariant lines $y=x+1$ and $y=x$. The local flow direction is colored red.}\label{cntFlow}
\end{figure}

\begin{figure}[htdp]
\includegraphics[width=1 \textwidth, height = 0.8\textwidth]{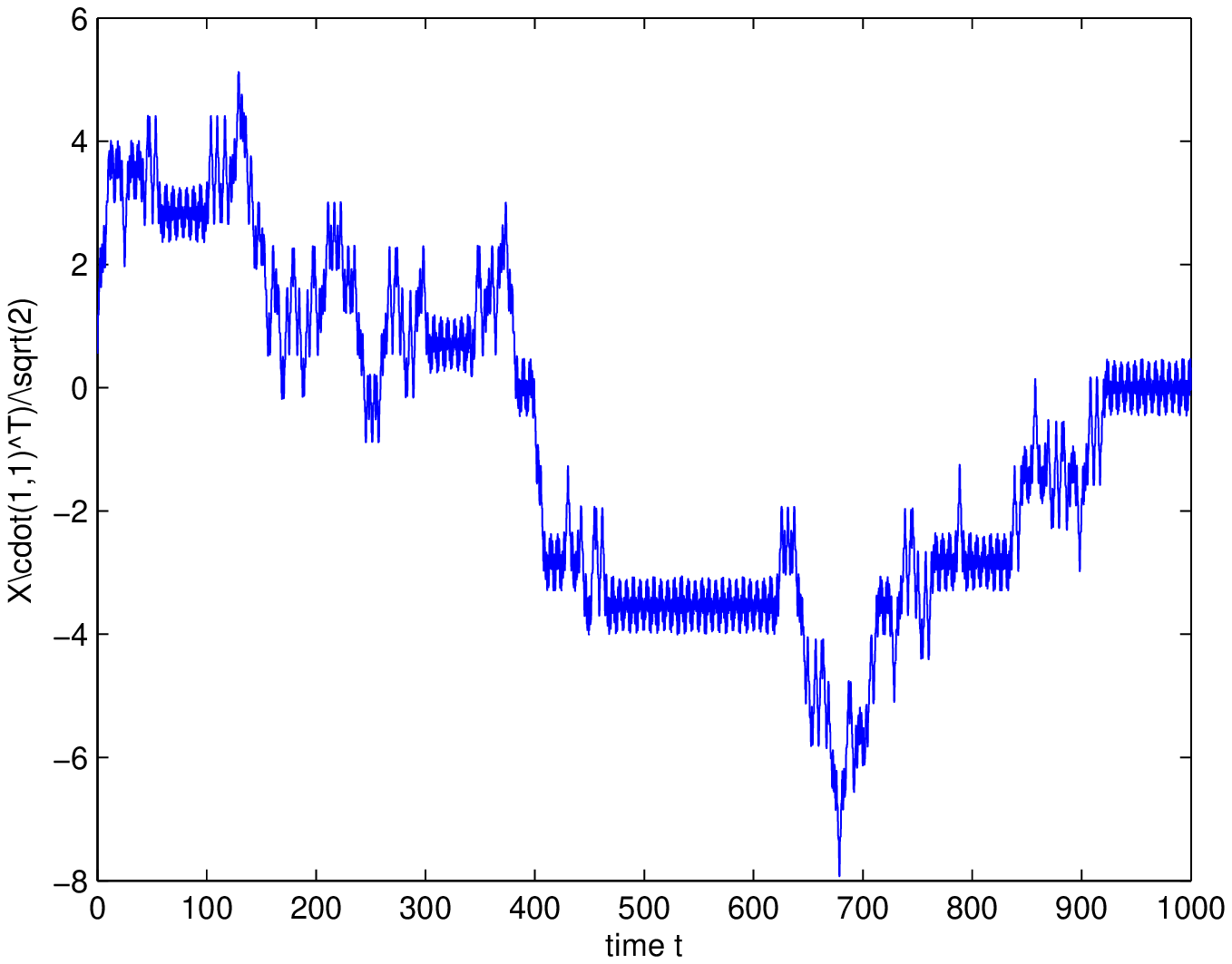} 
\caption{A projected trajectory $X(t)\cdot (1,1)/\sqrt{2}$ in time periodic cellular flow (\ref{unsteady}), 
$\theta =1$, showing stochastic character.}\label{projTr}
\end{figure}

For the unsteady flow 
\be
B_u=(\cos(2\pi y) + \theta \, \sin(2\pi y)\cos(t), \cos(2\pi x) + \theta \, \sin(2\pi x)\cos(t)),\label{unsteady}
\ee 
the invariant manifold consists of lines $y=x+n$, $n \in \mathbb{Z}$. 
The flow trajectories are restricted in the channels bounded by two neighboring lines 
$y=x+n+1$ and $y =x+n$. At $\theta > 0$, the flow trajectory extends itself from one cell square 
to another. The Lagrangian particle undergoes chaotic motion, see Fig.~\ref{cntFlow} for an 
illustration. The computation is done by a fourth order symplectic scheme.
The local flow direction is colored red. The intensified red region in the middle indicates 
that the particle spends a lot of time wandering in and out of the cells there. 
Fig.~\ref{projTr} plots a projected trajectory in the direction $(1,1)/\sqrt{2}$ vs. time. 
The stochastic feature is visible.  

\begin{figure}[ht!]
\includegraphics[width=0.5\textwidth, height = 0.45\textwidth]{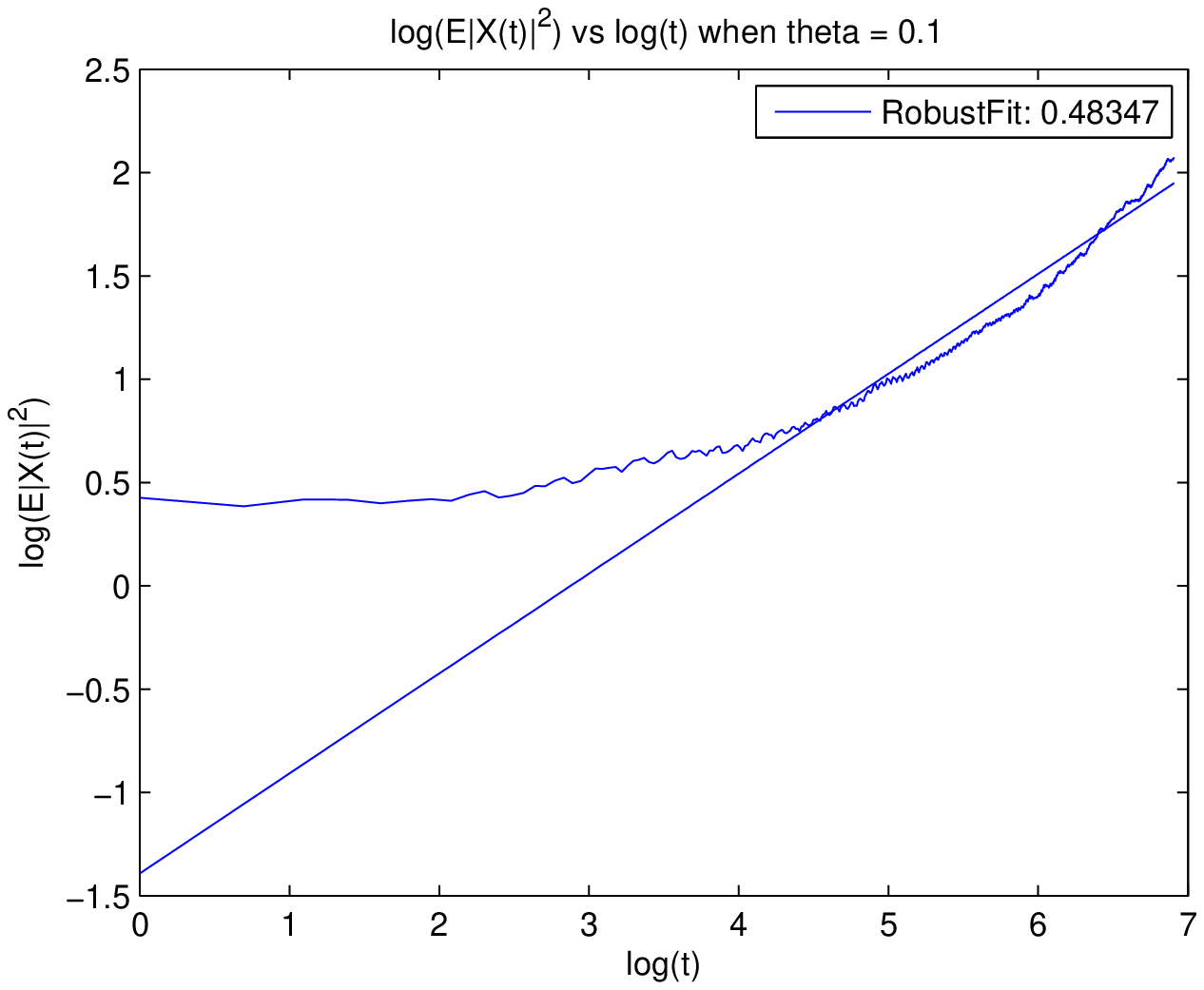}
\includegraphics[width=0.5\textwidth, height = 0.45\textwidth]{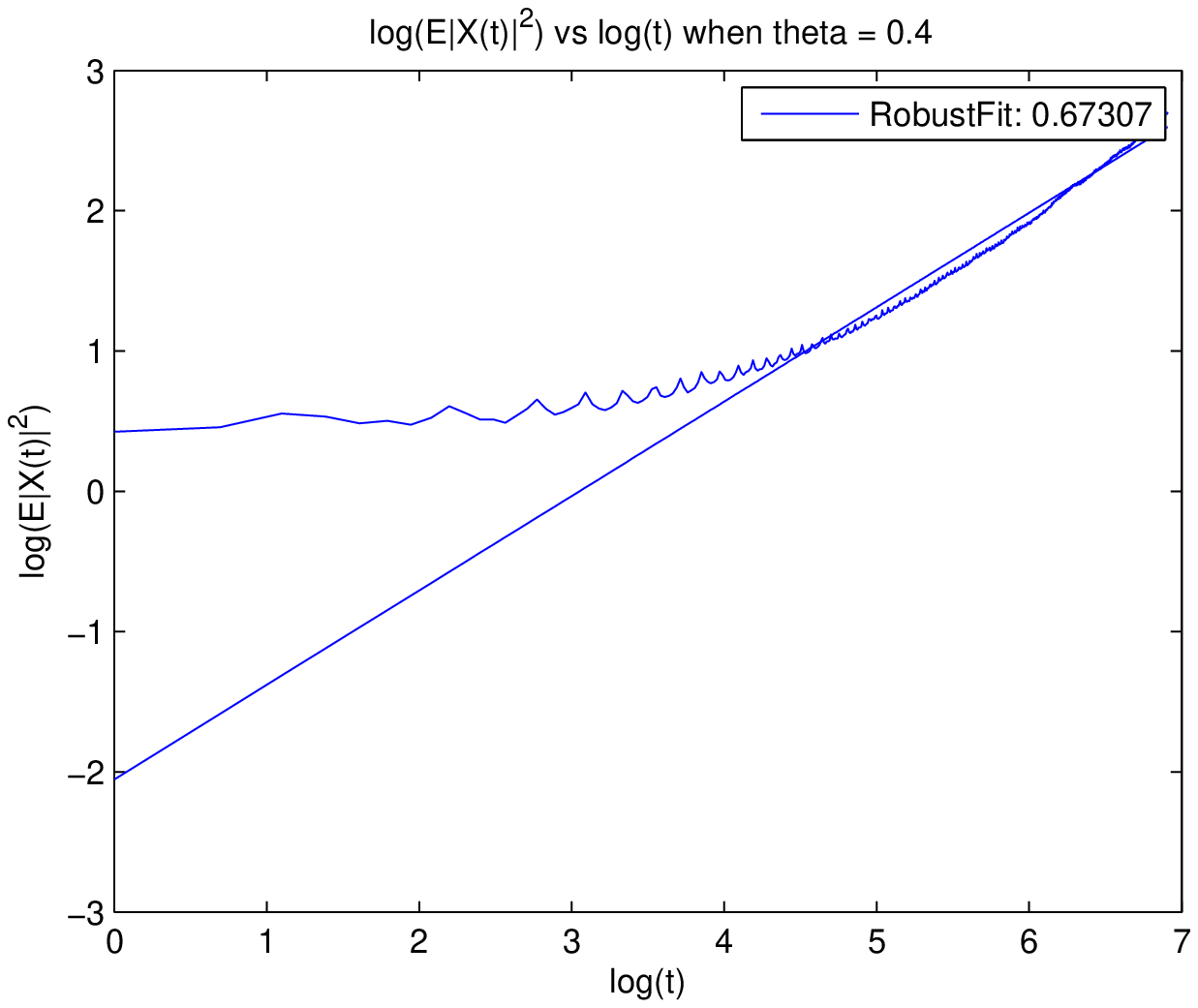}
\caption{Mean square distances $E[|X(t,\omega)|^2]$ vs. time on the logarithmic scale with robust fit showing sub-diffusive scaling $O(t^p)$,
$p=0.483, 0.673 $ at $\theta=0.1$ (left) and $0.4$ (right).}\label{subd1}
\end{figure}

\begin{figure}[ht!]
\includegraphics[width=0.5\textwidth, height = 0.45\textwidth]{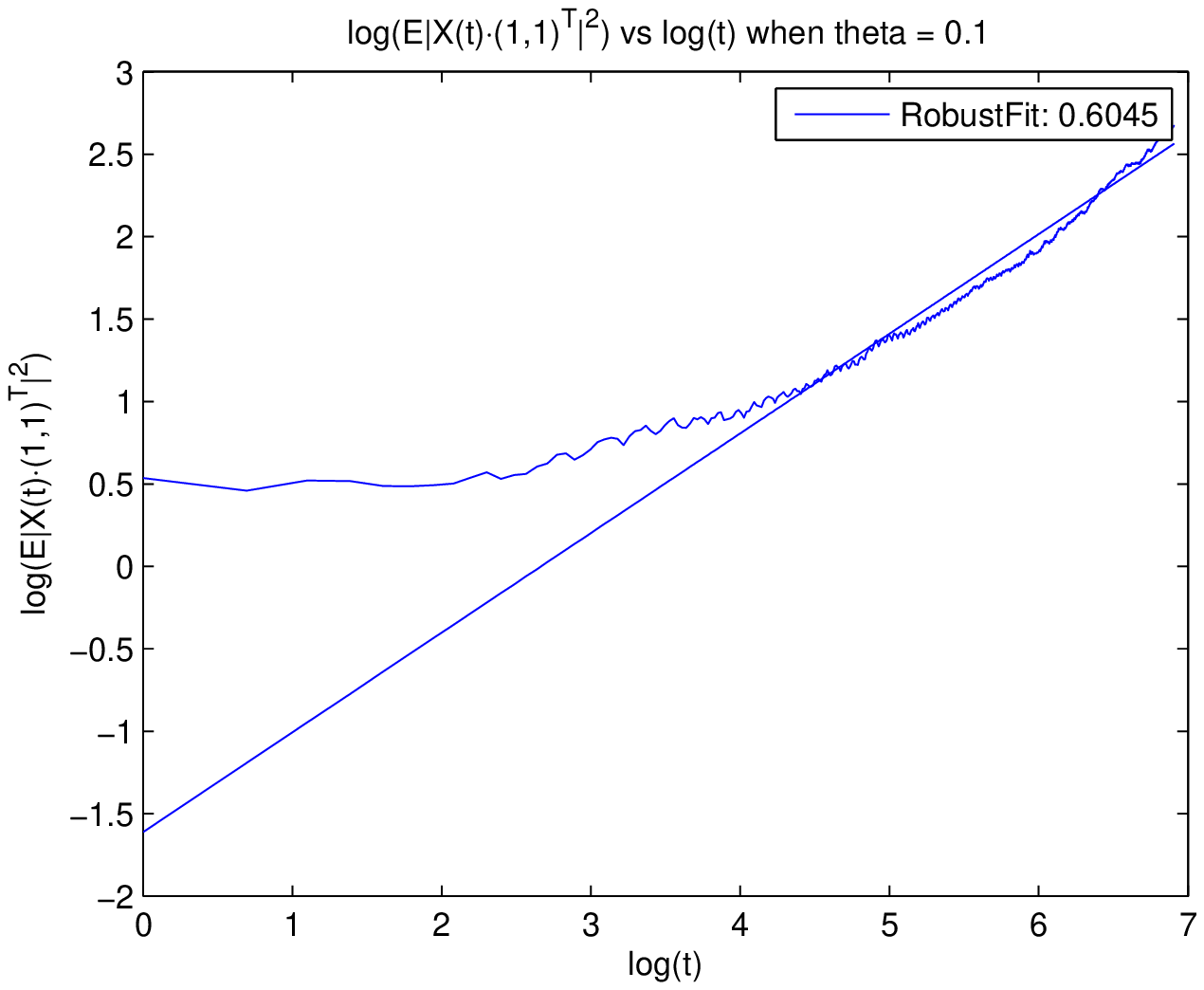}
\includegraphics[width=0.5\textwidth, height = 0.45\textwidth]{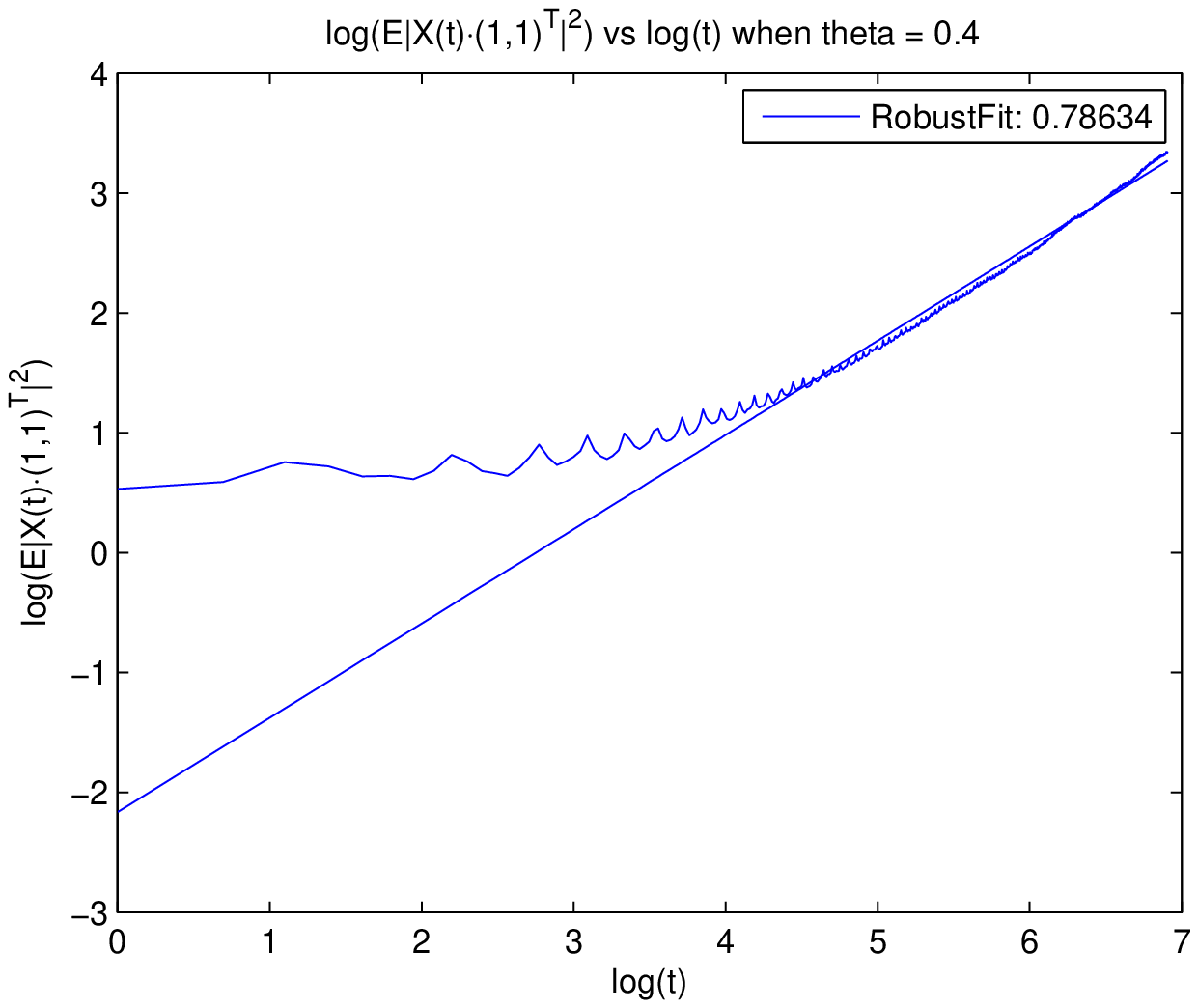}
\caption{Projected mean square distances $E[ (X(t,\omega)\cdot (1,1))^2 ]$ vs. time on the logarithmic scale with robust fit 
showing sub-diffusive scaling $O(t^q)$, $q=0.605, 0.786$ at $\theta=0.1$ (left) and $0.4$ (right).}\label{subd2}
\end{figure}

To quantify the disorder and draw a connection with diffusion process, 
we compute $10^4$ trajectories $X(t,\omega)$ over time interval $[0,1000]$ with uniformly distributed initial points on a base cell square, 
where $\omega$ denotes the random samples. We calculate the empirical mean square distance ($E[|X(t,\omega)|^2]$)
and the projected mean square distance 
$E[ (X(t,\omega)\cdot (1,1))^2 ]$, and plot them as a function of time on the logarithmic 
scale to recover scaling laws. For efficiency, the samples are obtained by a 500 node parallel implementation of 
the standard 4th order Runge-Kutta method. Fig.~\ref{subd1} shows that the mean square distances  
$E[|X(t,\omega)|^2] \sim O(t^p)$, $p\approx 0.483, 0.673$ at $\theta =0.1, 0.4$ respectively, hence belonging to the sub-diffusive regime. 
Similar sub-diffusive behavior can be observed in Fig.~\ref{subd2} where the projected mean square distance scales like  
$E[ (X(t,\omega)\cdot (1,1))^2 ]\sim O(t^q)$, $q \approx 0.605, 0.786$ at $\theta =0.1, 0.4$.

\section{Numerical Methods}
Numerical approximation of advection-dominated problems is a 
topic of independent interest. We shall use the 
Edge-Averaged Finite Element (EAFE) method~\cite{XZ99} due to the nice  
discrete maximum principle obeyed by EAFE and the corresponding fast multigrid solvers. 
On the other hand, pseudo-spectral methods are a class of highly accurate numerical methods for 
solving partial differential equations. 
In practice, pseudo-spectral method has excellent error reduction properties 
with the so-called ``exponential convergence'' being the fastest possible 
as long as the solution is smooth. 
In this section, we briefly review these two methods used in our simulation.


Recall that we are solving the following evolution 
problem over the unit square domain $\Omega = [0,1]\times[0,1]$ with 
periodic boundary condition and a constant initial condition:
\begin{equation}\label{eq:linear5}
\left\{
\begin{array}{l}
w_t = \epsilon\triangle w +(2\epsilon\lambda\vec{e} + B)\cdot\nabla w - (C_M - \epsilon\lambda^2 -\lambda\vec{e}\cdot B - \frac{1}{\tau}f'(0)) w,\\
w(0,y,t)=w(1,y,t); w(x,0,t) = w(x,1,t), \\
w(x,y,0)=1.
\end{array}
\right.
\end{equation}

\subsection{EAFE Method}
We first decompose the domain $\Omega$ into a triangulation $\mathcal T$ which is a set of triangles such that 
\begin{equation}
\cup_{\tau\in\mathcal{T}} \tau = \bar{\Omega}\quad \text{and}\quad \mathring{\tau}_i\cap\mathring{\tau}_j = \emptyset.
\end{equation} 
For the unit square, we simply use the uniform mesh obtained by setting length of each triangle $h$. 

We describe the EAFE method using a simple advection-diffusion equation
\begin{equation}\label{eq:CReq}
\left\{
\begin{array}{ll}
-\nabla\cdot(\nabla u + \beta(x) u) = f(x), &x\in\Omega\\
u=0 & x\in \partial\Omega.
\end{array}
\right.
\end{equation}
Associated with each $\mathcal{T}_h$, let $V_h\subset H^1_0(\Omega)$ be the piecewise 
linear finite element space. The space $H^1_0(\Omega)$ is defined as the subspace of $H^1(\Omega)$ 
with zero trace on $\partial\Omega$.

Given any edge $E$ in the triangulation, we introduce a function $\phi_E$ defined locally on $E$ 
(up to an arbitrary constant) by the relation:
\begin{equation}
\frac{\partial\phi_E}{\partial t_E} = \frac{1}{|t_E|} \epsilon^{-1}(\beta\cdot t_E),
\end{equation}
where $E$ is the edge connecting two vertices $q_i$ and $q_j$, and $t_E$ is the vector such that $t_E = q_i-q_j$.
	
The EAFE formulation of the problem (\ref{eq:CReq}) is: Find $u_h\in V_h$ such that
\begin{equation}
a_h(u_h,v_h) = f(v_h)\quad \text{for}\ \text{any}\ v_h\in V_h
\end{equation}
where
\begin{equation*}
a_h(u_h,v_h) = \sum_{T\in\mathcal{T}_h}
\left\{
\sum_{E\subset T} 
\frac{\omega^T_E}{|t_E|} \int_E e^{\phi_E}J(u)\cdot t_Eds
\ \delta_E v_h
\right\}
\end{equation*}
and
\begin{equation*}
\left\{
\begin{array}{l}
\omega_E = \frac{1}{2} \sum_{E\subset T}\cot\theta_E^T \geq 0\\
J(u) = \epsilon\nabla u + \beta u
\end{array}
\right.
\end{equation*}
where $\theta_E^T$ is 
the angle between the edges that share the common edge $E$ and $\delta_E$ is related to the tangential derivative along $E$.

The authors of~\cite{XZ99} show that if the triangulation is a so-called Delaunay triangulation, i.e., the summation of two angles of an interior edge is less than or equal to $\pi$, the matrix of the linear system generated by EAFE 
would be an M-matrix with nonnegative row sum. The detail of the scheme can be found in~\cite{XZ99}.

When applied to equation \eqref{eq:linear5}, notice that $\vec{B}$ is divergence free, the operator can be written in the form of \eqref{eq:CReq} and the reaction term is always positive. We apply the mass lumping method~\cite{TPB71} so that the discretization of the reaction term makes positive contribution to the diagonal. Therefore the M-matrix property and consequently the discrete maximum principle still holds. 

\begin{rem}
We have tested the Box method~\cite{BBFS}, 
EAFE method~\cite{XZ99} and streamline diffusion method~\cite{ESW} on the advection-dominated problem. 
The accuracy of EAFE scheme is better than Box method, but lower than the streamline diffusion method. 
We chose EAFE instead of streamline diffusion scheme because  
the matrix of linear system generated by EAFE is M-matrix. 
The M-matrix not only preserves the discrete maximum principle 
(so that the numerical solution stays between 0 and 1) but also benefits from the 
fast solvers.   
\end{rem}


To accelerate the speed of computation, Algebraic Multi-Grid (AMG) solver is
used to solve the linear algebraic equation arising from each time step. When the matrix is an M-matrix, the corresponding AMG solver is proven to be efficient. Specifically for the advection-diffusion equations, a multigrid method preserving the M-matrix property in coarse level is developed in \cite{KXZ03}. 
Among many AMG software packages, we use AGMG (aggregation-based algebraic multigrid method) package~\cite{N}.
%
The numerical test shows that AMG solver is $8$ times faster than the direct solver.

\subsection{Pseudo-Spectral Method with Semi-Explicit Scheme}
To describe the pseudo-spectral method, we consider the problem:
\ba
& & u_t = \epsilon\Delta u + \vec{B}(x,y,t)\cdot\nabla u + Cu \nonumber \\
& &  u(x,y,0)=1; u(0,y,t)=u(1,y,t), \\
&& u(x,0,t)=u(x,1,t).
\ea
The derivatives of solutions are $\nabla u = \mathcal{F}^{-1} (i\vec{k}\mathcal{F}(u))$ 
and $\triangle u = \mathcal{F}^{-1} ((i\vec{k})^2\mathcal{F}(u))$. Here $\mathcal{F}$ is the discrete Fourier transformation and $\vec{k}$ represents the wave numbers.

The pseudo-spectral scheme with semi-explicit scheme is that diffusion term uses implicit scheme and advection term uses half step lagged explicit scheme:
\begin{equation}\label{eq:semi}
\frac{\hat{U}^{n+1}-\hat{U}^{n}}{\Delta t} = -\epsilon\vec{k}^2 \hat{U}^{n+1} + \mathcal{F}(\vec{B}\cdot \mathcal{F}^{-1}(i\vec{k}\hat{U}^n)) + C \hat{U}^n
\end{equation}
Here the discretization on the spatial domain is $N\times N$ and $\hat{U}^{n}$ is the frequency of $u$ at time step $n$. 
Pseudo-spectral method has higher order accuracy than the finite element methods.  
However, its time step size $\Delta t$ depends on the small diffusion parameter $\epsilon$. Therefore it is more costly to reach the large time solution. 

\subsection{Stability Analysis}
The pseudo-spectral method is not so efficient 
when diffusion parameter $\epsilon$ goes to zero. Here 
we compare the upwinding finite element method and pseudo-spectral method. 
The analysis suggests a hybrid algorithm combining finite element method and pseudo-spectral method.

\subsubsection{Upwinding Finite element method}
The stability condition of EAFE method 
is equivalent to that of an upwinding scheme.  To simplify analysis, 
the advection term is taken as one-dimensional and the reaction term is ignored. 
The proto-type equation is
\begin{equation}\label{proto1}
u_t = \epsilon\, u_{xx} + b\, u_x
\end{equation}

The upwinding semi-explicit scheme is
\begin{equation}\label{eq:upwinding}
\left\{
\begin{array}{l}
\frac{U^{n+1}_j-U^n_j}{\Delta t} = \epsilon(\frac{U^n_{j+1}+U^n_{j-1} - 2U^n_j}{2} + \frac{U^{n+1}_{j+1}+U^{n+1}_{j-1} - 2U^{n+1}_j}{2})/\Delta x^2 + b\frac{U^n_{j+1}-U^n_j}{\Delta x}\quad\text{if}\  b\geq 0 \\
\frac{U^{n+1}_j-U^n_j}{\Delta t} = \epsilon(\frac{U^n_{j+1}+U^n_{j-1} - 2U^n_j}{2} + \frac{U^{n+1}_{j+1}+U^{n+1}_{j-1} - 2U^{n+1}_j}{2})/\Delta x^2 + b\frac{U^n_{j}-U^n_{j-1}}{\Delta x}\quad\text{if}\  b< 0
\end{array}
\right.
\end{equation}
Denoting $$ b^+=\max(b,0),\quad b^-=\min(b,0),$$ 
we carry out the von Neumann analysis to 
obtain the stability condition.
Upon substitution $U^n_j = \rho^n e^{ij\theta}$, (\ref{eq:upwinding}) becomes:
\begin{equation}\label{eq:upwinding2}
\frac{\rho-1}{\Delta t} = \frac{\epsilon(\rho+1)(\cos(\theta)-1)}{\Delta x^2} + \frac{|b|(\cos(\theta)-1+i\sin(\theta))}{\Delta x}
\end{equation}
Let $\mu = \frac{\epsilon\Delta t}{\Delta x^2}$ and 
CFL number $C = \frac{|b|\Delta t}{\Delta x}$. The 
Peclet number 
$$Pe = \frac{\text{advection}}{\text{diffusion}} = \frac{|b|\Delta x}{\epsilon} = C/\mu. $$
Equation (\ref{eq:upwinding2}) simplifies to:
\begin{equation}
\begin{array}{rl}
(1+\mu (1-\cos\theta))\rho =& (1-\mu(1-\cos\theta)) + Pe\, \mu(\cos\theta-1+i\sin\theta)\\
=& (1-\mu(1+\, Pe\, )(1-\cos\theta)) + i\, Pe\, \mu\sin\theta\\
\end{array}
\end{equation}

\noindent The stable condition is:
\begin{equation}
\rho \leq 1 
\Longleftrightarrow
 \mu \leq \frac{2(1-\cos\theta)(2+Pe) } {(1+Pe)^2(1-\cos\theta)^2 + (Pe\sin\theta)^2}
\end{equation}
It suffices to impose:
\begin{equation*}
\begin{array}{rl}
\mu\leq &\min_{\theta\in[-\pi,\pi]}\frac{2(1-\cos\theta)(2+Pe) } {(1+Pe)^2(1-\cos\theta)^2 + (Pe\sin\theta)^2}\\
=&(2+Pe)/Pe^2\\
\rightarrow& 1/Pe \quad\text{while}\quad \epsilon\rightarrow 0
\end{array}
\end{equation*}
The stability condition of the upwinding scheme in the limit of $\epsilon \downarrow 0$ is:
\begin{equation}
\Delta t\leq\frac{\Delta x}{|b|}, 
\end{equation}
a CFL condition independent of $\epsilon$.

\subsubsection{Pseudo-Spectral Method}\
Consider pseudo-spectral method on the proto-type equation (\ref{proto1}) again:
we have shown before:
\begin{equation}\label{eq:semispectral}
\frac{U^{n+1}-U^n}{\triangle t} = -\epsilon \vec{k}^2 U^{n+1} + b \mathcal{F}^{-1}(i\vec{k}\mathcal{F}U^n).
\end{equation}

To prove the stability of pseudo-spectral method with Semi-Explicit scheme, we assume:
\begin{equation}\label{eq:basis}
u(x,t) = \sum^{\infty}_{k=-\infty} \hat{u}_k(t) e^{ikx}.
\end{equation}
and we denote the projection operator:
\begin{equation}\label{eq:projection}
(P_N)u(x,t) = \sum_{-\frac{N}{2}<k\leq\frac{N}{2}} \hat{u}_k(t) e^{ikx}.
\end{equation}

If we plug (\ref{eq:basis}) and (\ref{eq:projection}) into (\ref{eq:semispectral}), we have that the k-th component of the vector, which should be the corresponding coefficient of $e^{ikx}$:
\begin{equation}\label{eq:elementwise}
\frac{\hat{u}^{n+1}_k-\hat{u}^{n}_k}{\triangle t} = -\epsilon k^2 u^{n+1}_{k}  + ik b \hat{u}^n_k.
\end{equation}
From \eqref{eq:elementwise}, we obtain
\begin{equation}
|\hat{u}^{n+1}_k| = \left |\frac{1+ikb\triangle t}{1+\epsilon k^2\triangle t} \hat{u}^n_k \right | \leq \frac{|1+ikb\triangle t|}{|1+\epsilon k^2 \triangle t|} |\hat{u}^n_k|
\end{equation}
So $U^n$ in (\ref{eq:semispectral}) is stable if and only if:
\begin{equation}
\left |\frac{1+ikb\triangle t}{1+\epsilon k^2 \triangle t}\right |\leq 1 \quad \forall k \in \{ -\frac{N}{2}<k\leq \frac{N}{2}\}
\end{equation}
which is equivalent to
\begin{equation}
\begin{array}{rll}
& \triangle t \leq \frac{2\epsilon}{b^2-\epsilon^2 k^2}& \forall k \in \{ -\frac{N}{2}<k\leq \frac{N}{2}, k\neq 0, b^2-\epsilon^2 k^2 >0 \}
\end{array}
\end{equation}
So if the equation is advection dominant, then at least $k=1$ satisfies  
\begin{equation}
-\frac{N}{2}<k\leq \frac{N}{2},\quad k\neq 0,\quad b^2-\epsilon^2 k^2 >0
\end{equation}
and the stable condition would be $\triangle t\leq  \frac{2\epsilon}{b^2-\epsilon^2}$. When $ \epsilon \ll 1 $, the time step is in the order of $\mathcal O(\epsilon)$ which is very restrictive.

\section{Numerical Results on Residual Speeds}
\begin{figure}
\includegraphics[width=0.5\textwidth, height = 0.4\textwidth]{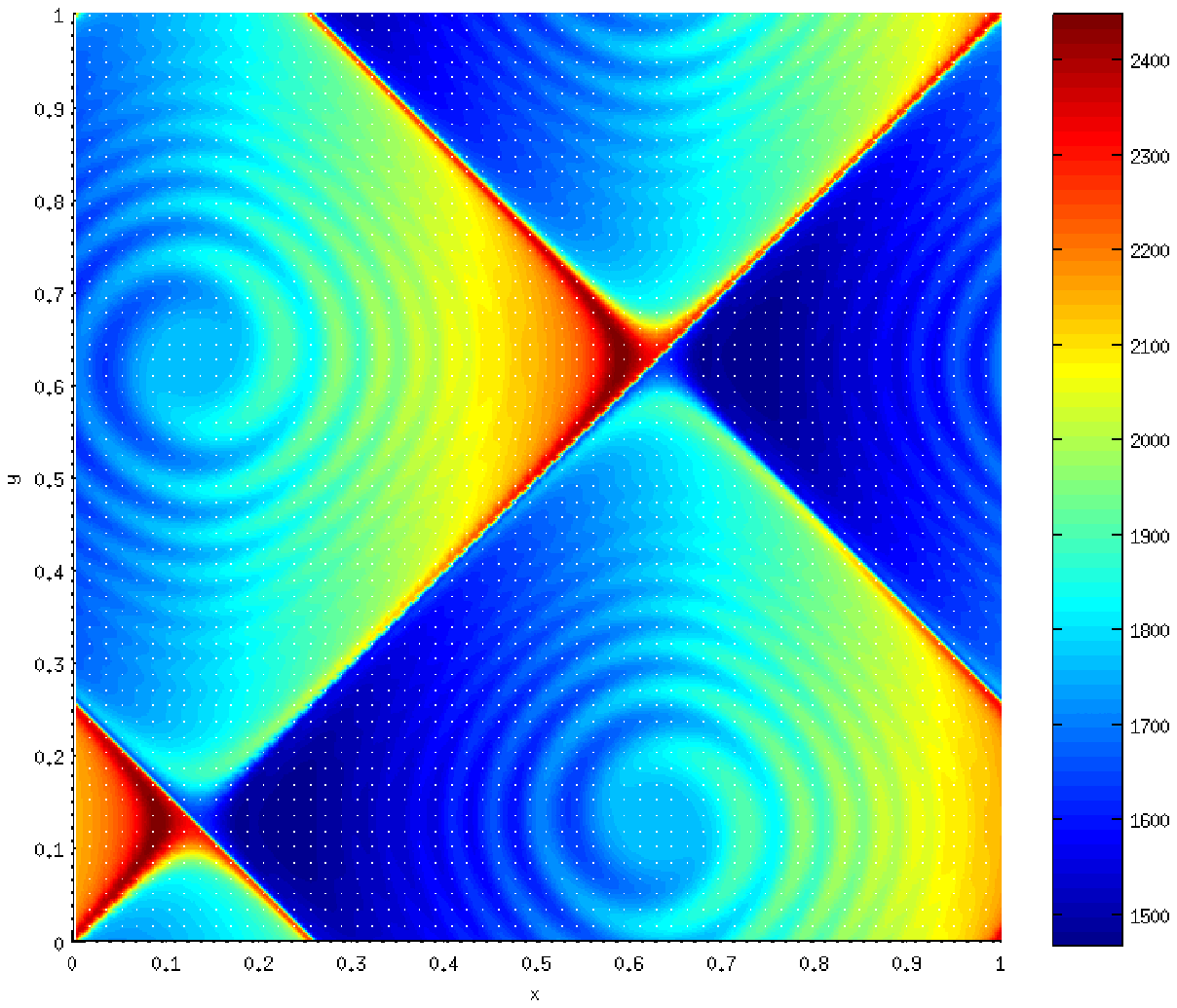}
\includegraphics[width=0.5\textwidth, height = 0.4\textwidth]{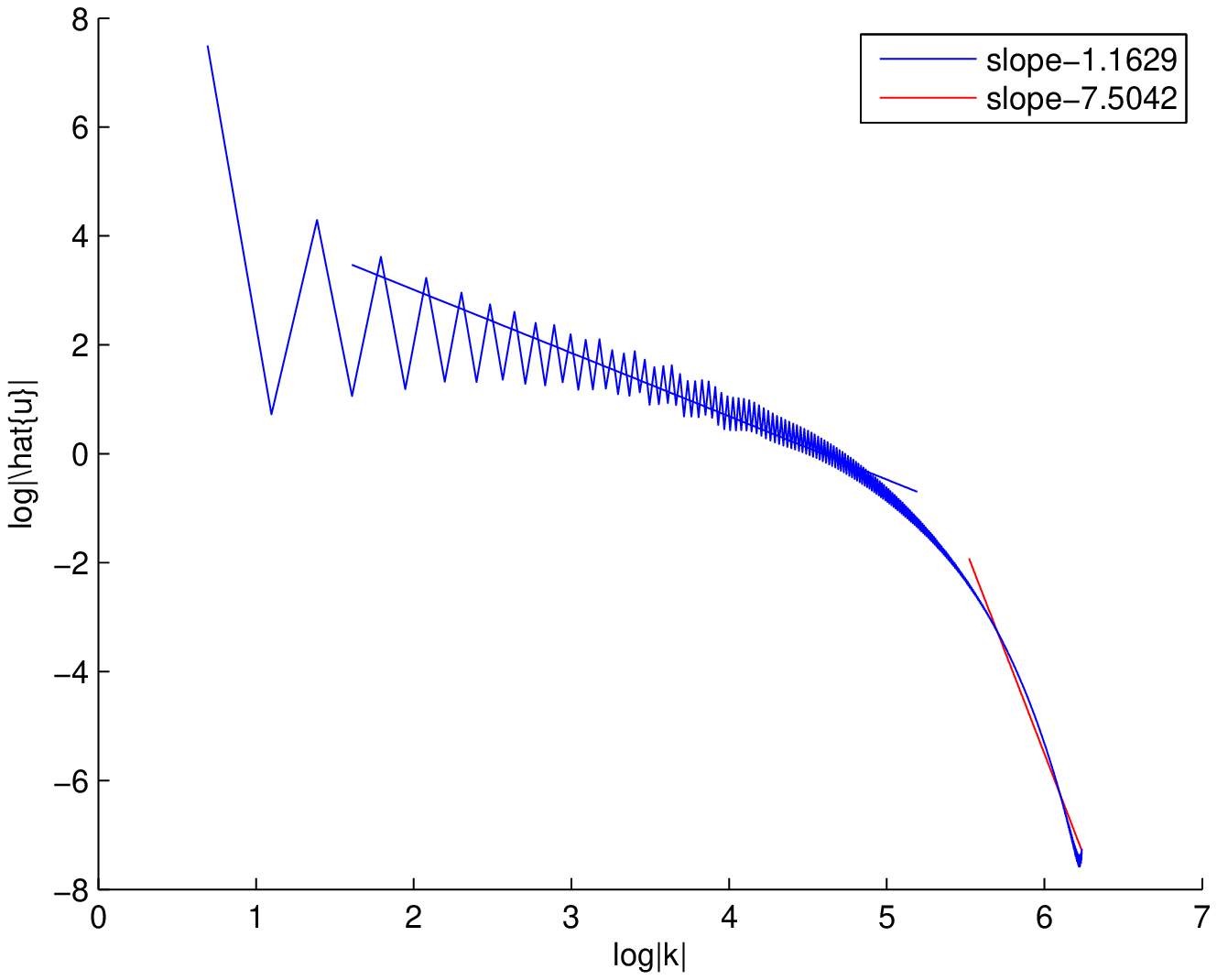}
\caption{A quasi-stationary solution (left) at time $t=2$, $\epsilon=10^{-3}$, $|B|=500$, and $\lambda=1$. 
The number of space nodes $N=2^9$ in each direction and time step $dt=1/10^8$. The power spectrum of the 
auxiliary field $w$ (right) for the unsteady flow~\eqref{unsteady} shows that 
the energy decreases towards high frequency, with a scaling range (slope = -1.1629) 
in the intermediate region of wave numbers.}\label{w_hat}
\end{figure}

For each given $\lambda$, we compute the Lyapunov exponent of~\eqref{varp1} by finding the long time 
quasi-stationary solution $u$ and applying formula \eqref{lyapunov}. A quasi-stationary 
solution with $\lambda=1$ is shown in the left plot of Fig.~\ref{w_hat}. The layer structure of the numerical solution 
moves along the $y=x$ direction and oscillates between the two neighboring invariant manifolds (lines of unit slope). 
The right plot of Fig.~\ref{w_hat} shows the power spectrum of the auxiliary field (a genearalized eigenfunction) 
$w$ (right) for the unsteady flow~\eqref{unsteady}. We observe that the energy decreases 
towards high frequency, with a linear decay (scaling range) in the intermediate region of wave numbers.

After calculating $\mu$ by the algorithm discussed in (\ref{lyapunov}), the minimization problem (\ref{eq:inf}) is solved 
by the golden-section method\cite{B73}. Golden-section method is a method to find the minimal point of a particular function called unimodal function. The idea of Golden-section method is by successively narrowing down the range of searching interval. A common definition of unimodal function is for some value $m$, it is monotonically decreasing when $x<m$ and monotonically increasing when $x\geq m$. 
In~\cite{NX05}, the authors show that the function $\frac{\mu(\lambda)}{\lambda}$ is indeed a unimodal function which is  
strictly convex with respect to $\lambda$, 
decreasing over interval $[0, \lambda^*]$ and strictly increasing 
over interval $[\lambda^*, \infty]$, with $\lambda^*$ the unique
minimal point of $\frac{\mu(\lambda)}{\lambda}$. Notice that Newton's method is 
not applicable since $\mu'(\lambda)$ is not known.

%

\begin{figure}[ht!]
\centering
\includegraphics[width=1\textwidth, height = 0.8\textwidth]{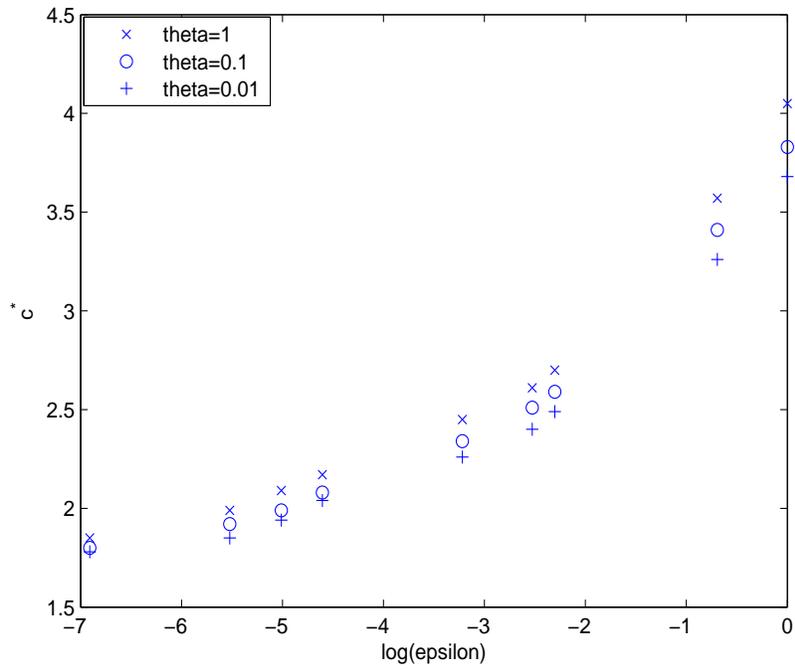}
\caption{Residual KPP front speeds in the small diffusion limit $\epsilon \downarrow 0$.}
\label{residual}	
\end{figure}

For time periodic cellular flow (\ref{tdcell}), EAFE method and spectral method are combined together to obtain the result in Fig.~\ref{residual}. The initial searching interval is $[0, 1000]$. EAFE is used first to narrow down the searching interval to an neighborhood of $\lambda^*$ with length 2. The space size and time step is $h = 1/2^{10}$ and $dt=1/10^7$, respectively. Since the minimal diffusion parameter $\epsilon=\mathcal{O}(10^{-3})$ and amplitude of advection term is $\mathcal{O}(10^{3})$, the choice of $dt = 1/10^7$ leads to numerical method stable. The spectral method with space node $N = 2^9$ in each direction and time step $dt=1/10^8$ is used to obtain the minimal point when the length of searching interval is less than $2$. The stopping criterion of searching algorithm is that the difference of two consecutive value of $\frac{\mu(\lambda)}{\lambda}$ is less than $5\times10^{-3}$. The benefit of using two methods together is that we can find the minimal point faster without the sacrifice  of the accuracy. 

\begin{figure}[ht!]
\centering
\includegraphics[width=1\textwidth, height = 0.8\textwidth]{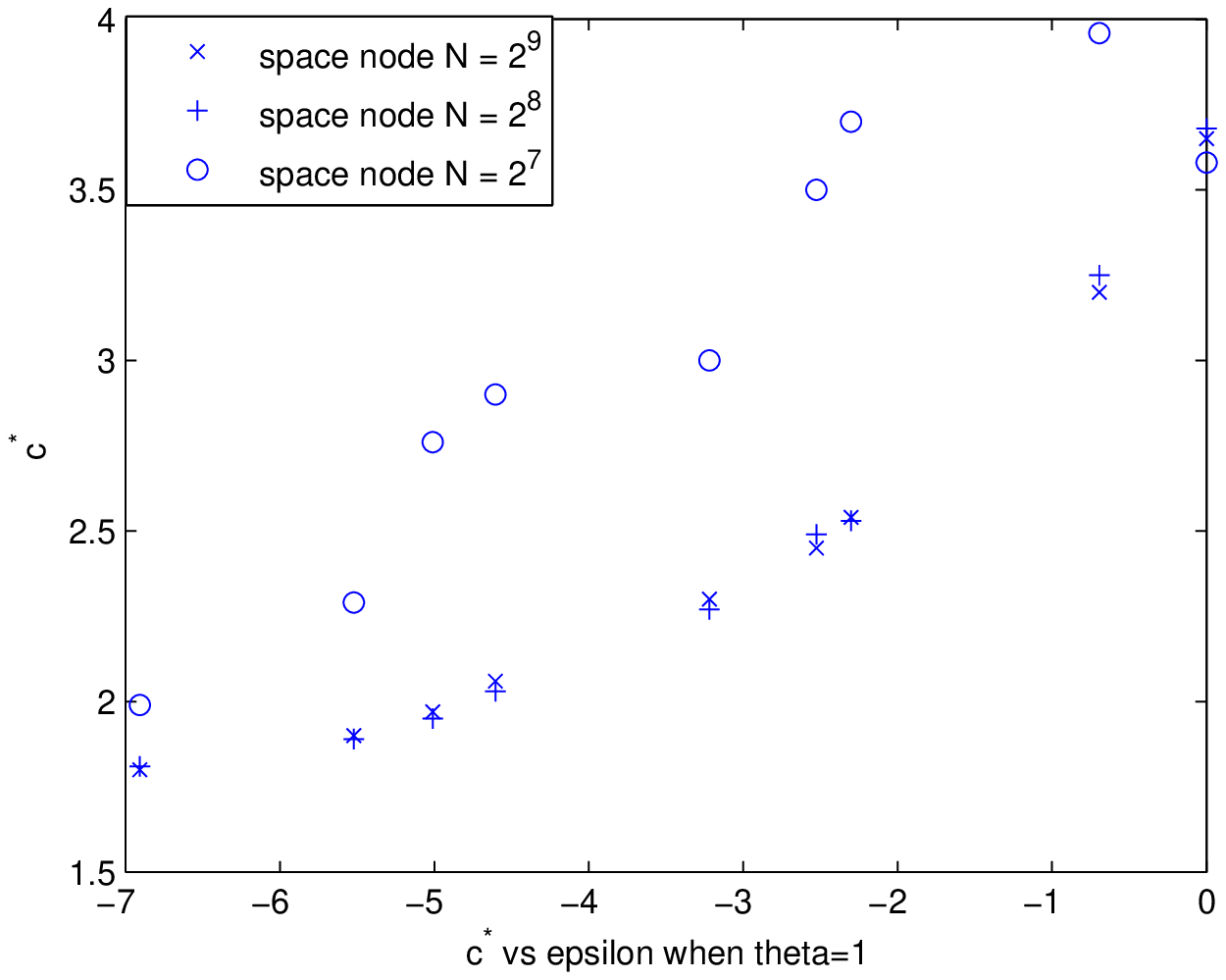}
\caption{Residual KPP front speeds at $\theta =1$ under grid refinement.}
\label{residualGrid}	
\end{figure}

In Fig.~\ref{residual}, we plot $c^*$ as a function of $\epsilon$ for 
three values of parameter $\theta =10^{-2},10^{-1},1$. Recall that $\theta$ is the perturbation parameter which determines how close the flow is to the steady cellular flow; see \eqref{tdcell}. 
From Fig.~\ref{residual}, we observe that for all three positive values of $\theta$, $c^*(\epsilon) = O(1)$ 
as $\epsilon \downarrow 0$. In contrast, it is well known that $c^{*}(\epsilon) = O(\epsilon^{1/4})$ in 
steady cellular flows\cite{NR05}. The presence of chaotic 
trajectories in time periodic cellular flows contributed to this phenomenon. 
They are much more mobile and far reaching than their counterparts of steady cellular flows.
Fig.~\ref{residualGrid} shows the persistence of residual front speeds 
under grid refinement at $\theta = 1$ and the close proximity of the data points 
between those at grid sizes of  $N = 2^8$ and $N = 2^{9}$ validates the convergence of the results numerically. 

\section{Concluding Remarks}
We have studied the KPP front speed asymptotics computationally 
in time periodic cellular flows with chaotic streamlines in the 
small molecular diffusivity limit. The chaotic streamlines statistically resemble a sub-diffusion process and 
enhance the KPP front speed $c^*$ significantly in the sense that $c^*$ has a positive limit 
as molecular diffusivity tends to zero. Such residual transport phenomenon is absent in steady 
cellular flows with ordered streamlines. To facilitate effective computation in the advection dominated regime, 
we combined an upwinding finite element methods with the spectral method for computing the principal eigenvalue 
of a time periodic parabolic operator with small diffusion and for a subsequent minimization. 
In future work, we plan to study residual KPP front speeds in three space dimensional chaotic flows.

\section{Acknowledgements}
The work was partially supported by NSF grant DMS-1211179. We thank Tyler McMillen for 
helpful conversations and visualization of chaotic dynamical systems. L.C 
was also supported by NSF grant DMS-1115961 and DOE prime award \# DE-SC0006903.

\end{document}